\documentclass[onecolumn,12pt]{article}
\usepackage{wrapfig}
\usepackage{graphicx,latexsym}
\usepackage{amssymb,amsthm,amsmath,amsbsy}
\usepackage[position=top]{subfig}
\usepackage[]{abstract}
\usepackage{mciteplus}
\usepackage{achemso}
\usepackage{bm}  
\usepackage[]{authblk}
\usepackage[stable,symbol]{footmisc}
\usepackage[usenames]{color}
\usepackage{setspace}
\usepackage[implicit=false]{hyperref}

\newcommand\single{%
  \renewcommand\baselinestretch{1.0}%
  \renewcommand\arraystretch{1.0}%
  \normalsize%
}

\newcommand{\eb}{\varepsilon_\mathrm{b}}
\newcommand{\degree}{^\circ}

\newcommand{\scL}{\mathcal{L}}
\newcommand{\es}{\varepsilon_\mathrm{s}}
\newcommand{\esph}{\es}
\newcommand{\eaa}{\varepsilon_\mathrm{AA}}
\newcommand{\nd}{N\Delta34}
\newcommand{\xs}{\chi_\mathrm{s}}
\newcommand {\apgt} {\ {\raise-.5ex\hbox{$\buildrel>\over\sim$}}\ }
\newcommand {\aplt} {\ {\raise-.5ex\hbox{$\buildrel<\over\sim$}}\ }
\newcommand{\lessim}{\aplt}
\newcommand{\greatsim}{\apgt}
\newcommand{\csurf}{c_\mathrm{surf}}
\newcommand{\bb}{\mathbf{b}}
\newcommand{\rc}{r_\mathrm{c}}
\newcommand{\rs}{R_\mathrm{s}}
\newcommand{\tc}{\theta_\mathrm{c}}
\newcommand{\pc}{\phi_\mathrm{c}}
\newcommand{\kt}{k_\mathrm{B}T}
\newcommand{\gamc}{\gamma_\mathrm{c}}
\newcommand{\ftone}{f_{\mathrm{T}1}}
\newcommand{\nmorph}{n_\mathrm{c}}
\newcommand{\sect}[1]{  { \noindent \flushleft \bf \large #1} \noindent}
\newcommand{\ssect}[1]{ {\noindent \flushleft \it #1}  }
\newcommand{\bssect}[1]{{\noindent \flushleft \bf #1}  }
\newcommand{\change}[2]{{#2}}
\newcommand{\Rij}{\mathbf{R}_{ij}}
\definecolor{Blue}{rgb}{0,0.0,1.0}
\definecolor{Red}{rgb}{1.0,0.0,0.0}

\title {Mechanisms of Size Control and Polymorphism in Viral Capsid Assembly}
\author{Oren M. Elrad \& Michael F. Hagan\footnote{{To whom correspondence should be addressed: hagan@brandeis.edu}}}
\affil{Department of Physics, Brandeis University\\ Waltham, MA 02454}

\begin{document}

\maketitle
\begin{abstract}

We simulate the assembly dynamics of icosahedral capsids from subunits that interconvert between different conformations \change{}{(or quasi-equivalent states)}.  The simulations identify mechanisms by which subunits form empty capsids with only one morphology, but adaptively assemble into different icosahedral morphologies around nanoparticle cargoes with varying sizes, as seen in recent experiments with brome mosaic virus (BMV) capsid proteins.  Adaptive cargo encapsidation requires moderate cargo-subunit interaction strengths; stronger interactions frustrate assembly by stabilizing intermediates with incommensurate curvature.  We compare simulation results to experiments with cowpea chlorotic mottle virus empty capsids and BMV capsids assembled on functionalized nanoparticles, and suggest new cargo encapsidation experiments. \change{}{Finally, we find that both empty and templated capsids maintain the precise spatial ordering of subunit conformations seen in the crystal structure even if interactions that preserve this arrangement are favored by as little as the thermal energy, consistent with experimental observations that different subunit conformations are highly similar.}

\end{abstract}

\sect{Introduction}

\noindent During the replication of many viruses, hundreds to thousands of protein subunits assemble around the viral nucleic acid to form a protein shell called a capsid.  In vitro studies show that capsid proteins can form particular empty capsid structures with high fidelity\cite{Johnson2005, *Casini2004, *Singh2003, *Willits2003, *Zlotnick2000}; yet capsids adopt different morphologies when challenged with nucleic acids\cite{Johnson2004, *Krol1999} or other cargoes\cite{Sun2007,*Dixit2006, *Chen2005, Chang2008, Hu2008} with sizes that are not commensurate with the preferred capsid structure.  No proposed dynamical mechanism simultaneously explains precise assembly of empty capsids and adaptable encapsidation of cargoes.  Understanding how viral components selectively assemble into the structure required for infectivity could spur the development of antiviral therapies that block or alter assembly.  At the same time, engineered structures in which viral capsids assemble around synthetic cargoes show great promise as delivery vehicles with adaptable sizes for drugs or imaging agents\cite{Soto2006, *Sapsford2006, *Boldogkoi2004, *Gupta2005, *Garcea2004, *Dietz2004}, and as subunits or templates for the synthesis of nanomaterials with exquisitely controlled sizes and morphologies\cite{Chatterji2005, *Falkner2005, *Flynn2003, *Douglas1998}.   Realizing these goals, however, requires understanding how properties of cargoes and capsid proteins dynamically determine the size and morphology of an assembled structure to enable adaptable assembly.

In this work, we explore the interplay between cargo size and the morphology of icosahedral capsids with coarse-grained models that describe both the dynamic encapsidation of functionalized nanoparticles and the assembly of empty capsids.  Through our simulations, we uncover a mechanism by which subunits faithfully assemble into empty capsids with a single icosahedral morphology,  but also reproducibly assemble into different morphologies around nanoparticles with varying sizes, as seen in recent experiments \cite{Sun2007}. The model predicts that adaptability to cargo size is nonmonotonic with respect to the strength of subunit-cargo interactions.  This prediction can be tested in nanoparticle-capsid assembly experiments by varying the functionalized surface charge density on nanoparticles \cite{Dragnea2008}.
\\
\ssect{ Assembly of icosahedral viruses.} While at most 60 identical subunits can be arranged with icosahedral symmetry, Caspar and Klug showed that multiples of 60 proteins can form icosahedral capsids, if individual proteins take slightly different, or quasi-equivalent, conformations\cite{Johnson1997,*Zlotnick2005,*Caspar1962}. \change{}{These quasiequivalent conformations break the local 3-fold symmetry of the icosahedral face but, by assembling with precise spatial ordering of conformations, preserve the global icosahedral symmetry of the capsid. Despite their different geometry, the proteins interact with each other by interfaces that are substantially similar across different conformations.} A complete capsid is comprised of $60T$ subunits, where $T$ is the number of distinct protein conformations (see \ref{models}).

Although recent experiments\cite{Stockley2007} have begun to characterize subunit conformations during assembly and equilibrium theories have led to important extensions of quasi-equivalence \cite{Bruinsma2003, *Zandi2004, *Keef2005, *Chen2007, *Zandi2008, *Mannige2008},  the process by which the appropriate quasi-equivalent conformations are chosen during assembly remains poorly understood.  Berger and coworkers\cite{Berger1994} showed that Caspar-Klug structures result if assembly follows ``local rules'', in which only subunits with the \change{quasi-equivalent}{} conformation dictated by adjacent subunits can bind to an assembling capsid\change{, and Nguyen and coworkers\cite{Nguyen2008} recently simulated the assembly of $T$=3 capsids from subunits for which attractive interactions follow strict quasi-equivalence}{}.  There are two experimental observations, however,  that seem difficult to rationalize with conformation-dependent interactions. (1) How can subunit-subunit binding be conformation-specific for viruses in which subunit binding interfaces show little variation between conformations in capsid crystal structures (e.g. see \ref{hinge} and Ref.~\cite{Tang2006})?   (2) How can subunits that assemble with conformational specificity adapt to form capsids with different icosahedral morphologies around commensurate cargoes?  For example, Dragnea and coworkers \cite{Sun2007, *Dixit2006, *Chen2005, Dragnea2003} have demonstrated that brome mosaic virus (BMV) proteins assemble into  $T$=1, pseudo-T2, and $T$=3  capsids around \change{PEG-}{} functionalized nanoparticle cores with different diameters \change{}{that are functionalized with carboxylated polyethylene glycol.} 

We explore both of these questions here with a model for the assembly of $T$=1 and $T$=3  capsids from subunits that can interconvert between different conformations, with which we simulate the spontaneous assembly of empty capsids and the encapsidation of nanoparticles that template assembly of $T$=1 or $T$=3  capsids. \change{We systematically vary the extent to which binding between subunits favors quasi-equivalence, as well as an intrinsic bias for subunits to adopt particular conformations, suggested by recent experiments \cite{Tang2006}.  We find that a relatively weak preference for quasi-equivalent subunit binding enables robust assembly, while the intrinsic bias can control which morphology is favored.}{By systematically varying the extent to which binding between subunits depends on conformation, we show that even a weak conformational dependence ($\approx \kt$) enables robust assembly. In addition, we find that an intrinsic bias for subunits to adopt particular conformations, as suggested by recent experiments\cite{Tang2006}, can control which icosahedral morphology is favored.} We show that requiring the model to reproduce experimental observations for both empty and full capsids places tight constraints on model parameters, and thereby enables insights about morphology control in both systems.  In particular, we find a narrow, but physically reasonable, range of parameters for which only $T$=3 empty capsids assemble, but $T$=1 capsids form on a commensurate nanoparticle.
\newpage

\sect{Model}

\noindent We extend a class of models for $T$=1 capsids \cite{Hagan2006, Hagan2008} \change{}{(see Refs.~\cite{Nguyen2007,*Schwartz1998, *Hicks2006, *Wilber2007} for related models)}, in which subunits have spherically symmetric excluded volume and short-ranged, directional attractions between complementary interfaces. These interfaces are represented as `bond-vectors'  that rotate rigidly within the subunit frame of reference; we design their geometry such that the subunits tile an icosahedron according to \change{quasi-equivalence rules}{capsid crystal structures}. Thus, the lowest energy collective configurations correspond to ``capsids'' with $60T$ monomers in a shell with icosahedral symmetry. 

We focus on models for $T$=1 and $T$=3  capsids, for which the bond vector geometries are based on crystal structures of $T$=1 \cite{Larson2005} and  $T$=3  \cite{Reddy2001} BMV capsids.  Each subunit represents a protein dimer, the basic assembly unit for BMV \cite{adolph1974}.  The relative positions and conformations of subunits in a capsid are determined by associating each two-fold or quasi-two-fold dimer interface with a subunit center, as depicted in \ref{models}.  The orientations of bond-vectors and their complementarity are then determined from the relative locations of neighboring dimers; as shown in \ref{models} each interface between neighboring subunits is associated with a pair of complementary bond vectors. The resulting internal coordinates and list of complementary bond vectors are specified in the SI.

\begin{figure}[!h]
\begin{center}
\subfloat[]{
\includegraphics[width=0.50 \textwidth]{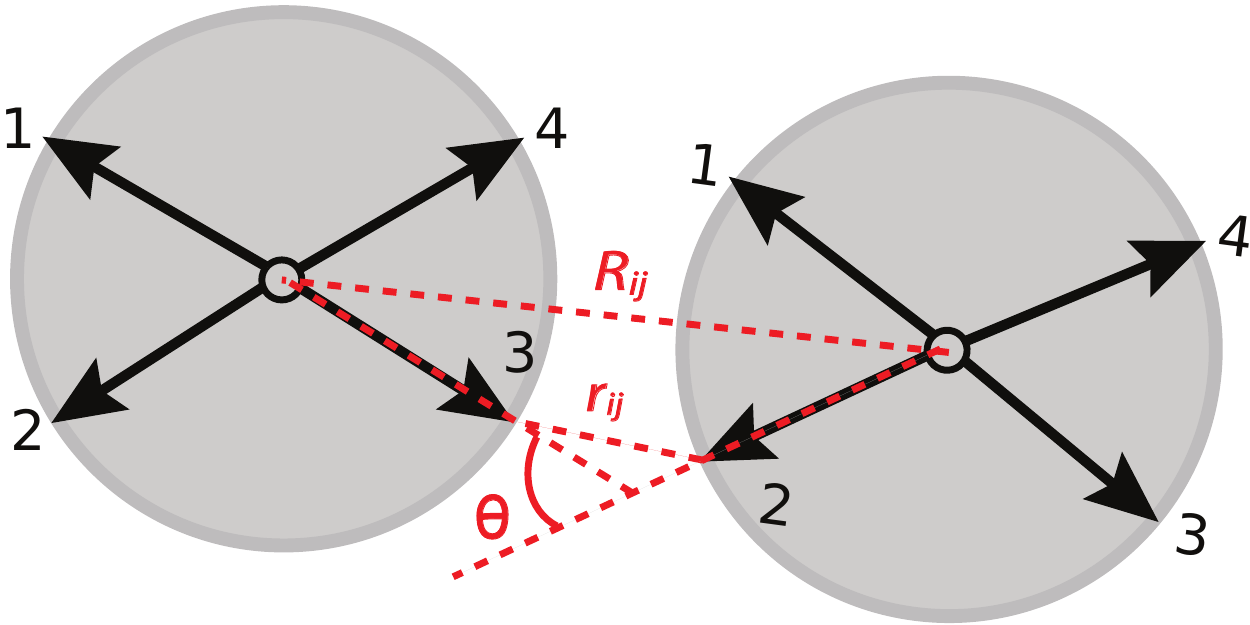}
\label{blah}
}
\subfloat[]{
\begin{tabular}{l}
\\
\\
$\Rij = \mathbf{R}_\text{i} - \mathbf{R}_\text{j}$ \\
\vspace{6pt}
$r_{ij}^b = | \Rij + \bb_j-\bb_i |$ \\
\vspace{6pt}
$\theta_{ij}^b=\arccos (-\bb_i \cdot \bb_j/|\bb_i||\bb_j|)$
\end{tabular}
}
\caption{
Definition of parameters used to calculate the interaction energy between complementary bond vectors (2) and (3). The dihedral angle $\phi_{ij}^b$ giving the rotation around the plane perpendicular to $R_{ij}$ is not shown. The bond vectors here are shown projected onto a plane even though the subunit geometries have intrinsic curvature (see \ref{models}).
}
\label{params}
\end{center}
\end{figure}

\begin{figure}
\begin{center}
\subfloat[T1]{
\includegraphics[width=0.45 \textwidth]{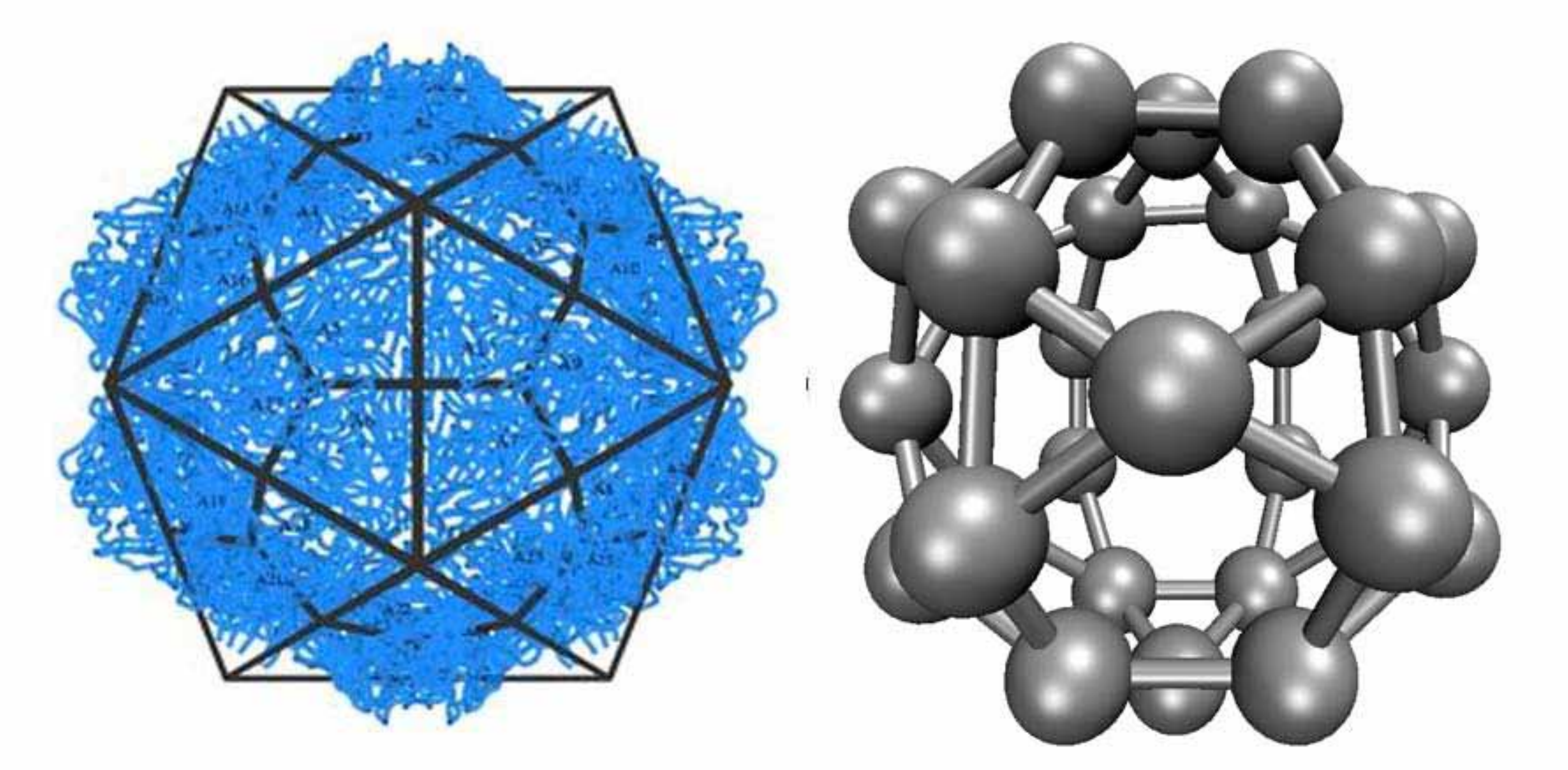}
\label{sublabel}
}
\subfloat[T3]{
\includegraphics[width=0.45 \textwidth]{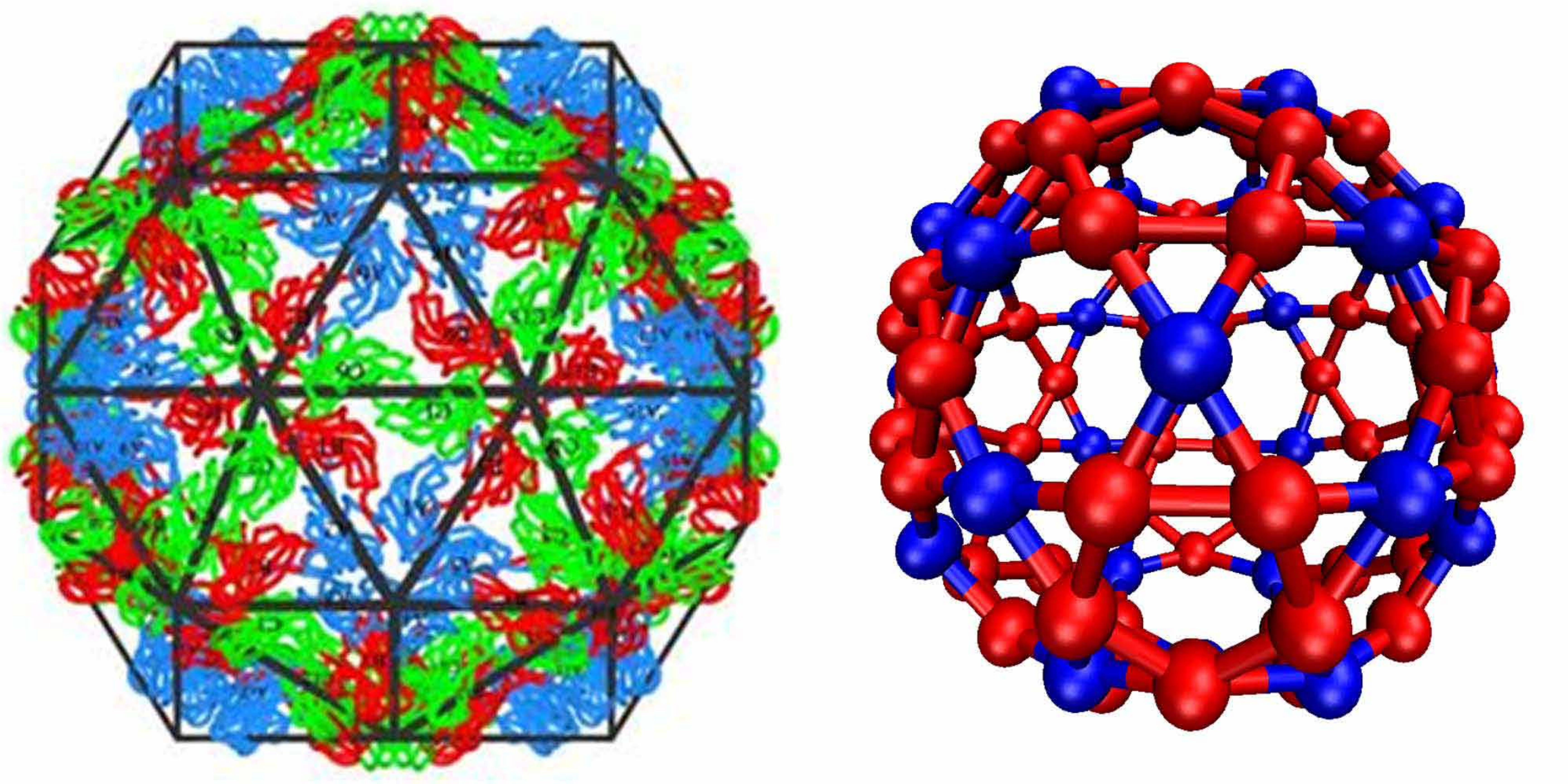}
}
\caption{
{\bf (a)} Images of brome mosaic virus (BMV) capsid crystal structures ($T$=1 \protect\cite{Larson2005},  $T$=3 , \protect\cite{Reddy2001} )overlaid with icosahedral cages. Subunits with quasi-equivalent conformations A, B, and C are blue, red, and green, respectively.  {\bf (b)} The model capsid geometries.  Each subunit represents a protein dimer, subunit sizes are reduced to aid visibility, and colors indicate the conformations of proteins within a dimer subunit: silver, AA; red, AB; blue, CC.}
\label{models}
\end{center}
\end{figure}

\begin{figure}
\begin{center}
\subfloat[]{
\includegraphics[width=0.55 \textwidth]{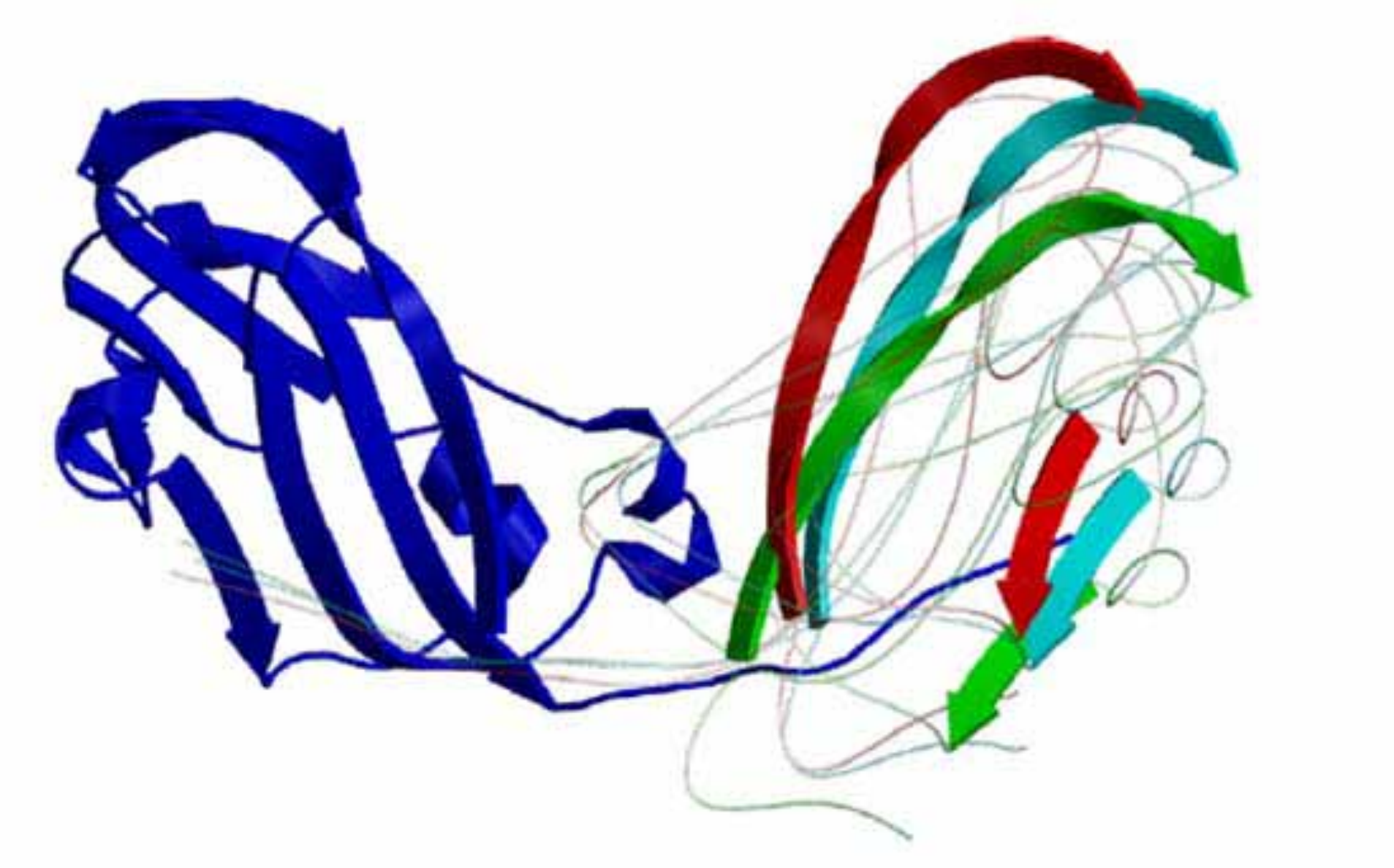}
}
\subfloat[] {
\includegraphics[width=0.4 \textwidth]{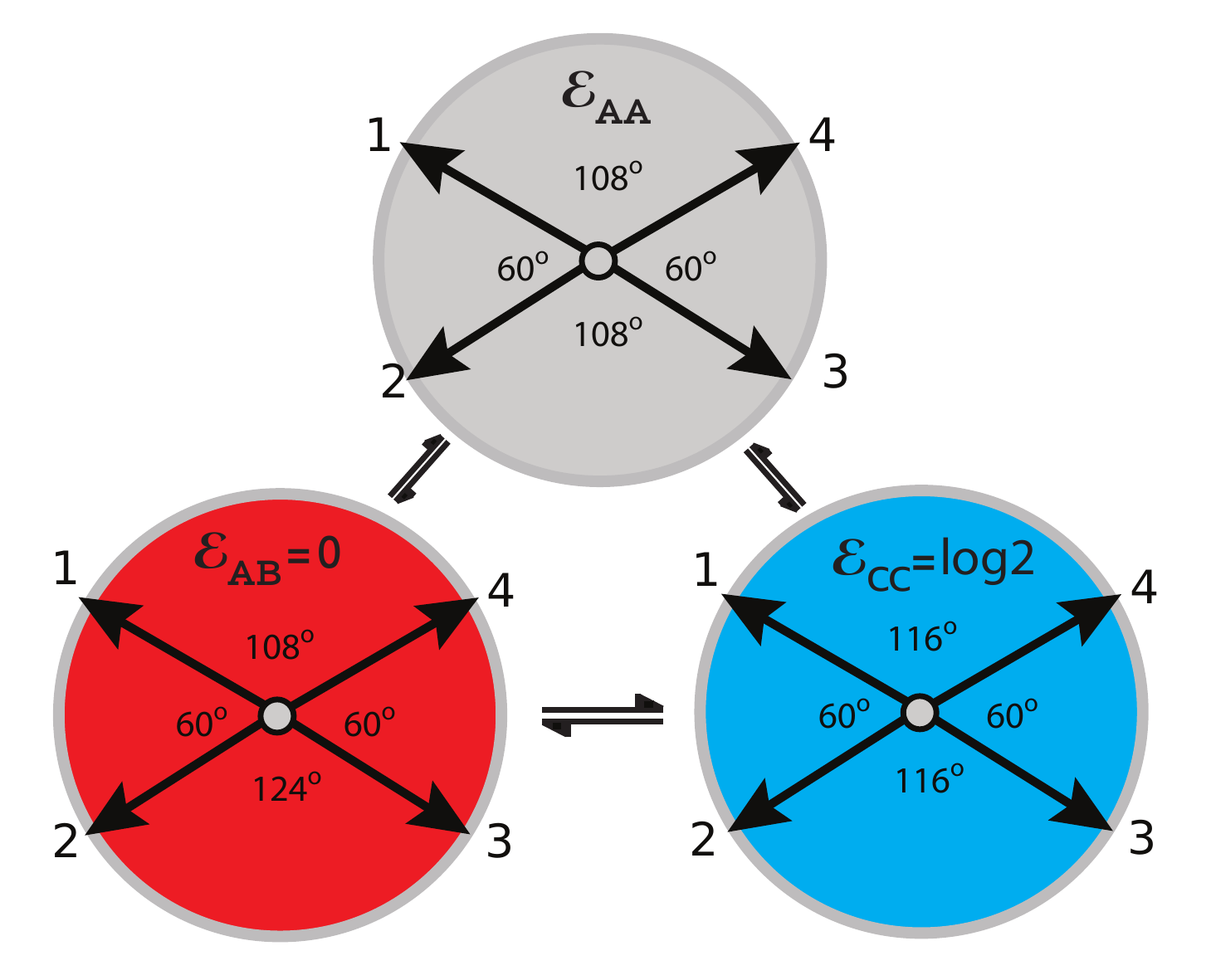}
}
\caption{{\bf (a)} Subunits with different conformations in CCMV capsids have nearly identical structures, but different angles of rotation about the ``dimer hinge'', which is the axis through the center of the dimer interface \protect\cite{Tang2006}.  One subunit is shown on the left (dark blue), and subunit positions on the right are indicated by the ribbon representations of two $\beta$-strands to show the range of rotation about the hinge for AB (red, $38\degree $), CC (light blue, $42\degree $), and AA (green, $45\degree $).   {\bf (b)} Geometries and intrinsic free energies (or hinge energies) for model subunits in different conformations. \change{}{The only geometrical difference between AA and CC subunits is the ``hinge'' angle between the left and right pairs of bond vectors -- the AA dimer has higher curvature (into the page) than the CC.} Bond vectors are depicted with arrows and the angles between them are indicated in degrees -- the angles do not add up to $360\degree$ because the vectors do not lie in a plane.}
\label{hinge}
\end{center}
\end{figure}

\ssect { Pair interaction.} The attractive interaction between two complementary bond-vectors $\bb_i$ and $\bb_j$ (see \ref{params}) is minimized when (1) the distance between the bond-vectors $r_{ij}^b$ is minimized,  (2) the angle between them, $\theta_{ij}^b$, is minimized and (3) the dihedral angle, $\phi_{ij}^b$, calculated from two \emph{secondary} bond-vectors which are not involved in the primary interaction (see the SI and \cite{Hagan2008}), is minimized. Requirement (3) creates an interaction that resists torsion and therefore enforces angular specificity in the plane perpendicular to the bond vector. The potential is given by equations (\ref{potential}) through (\ref{potential2})

\begin{eqnarray}
\label{potential}
U &=& U_{rep} + \sum_b U_{att}(r_{ij}^b)S(\theta_{ij}^b, \phi_{ij}^b) \\
U_{rep} &=& \Theta(R-2^\frac16)\ (\scL(R) + 1) \\
U_{att} &=& \Theta(r_{ij} - r_c) \eb \chi_{ij}^b (\scL(r_{ij} + 2^\frac16) - \scL(r_c))  \label{Uatt} \\ \nonumber
S(\theta,\phi) &=& \frac14 \ \Theta(\theta - \theta_c) \Theta(\phi - \phi_c)  \\
        &\phantom{=}&\left( \cos( \pi \theta / \theta_c) + 1\right) \left(\cos(\pi \phi / \phi_c) + 1 \right) \\
\scL(x) &\equiv& 4 (x^{-12} - x^{-6})
\label{potential2}
\end{eqnarray}
where the index $b$ sums over pairs of complementary bond vectors, $\Theta(x)$ is the Heaviside step function,  $R$ is the subunit center-to-center distance, $\scL(x)$ is a `Lennard-Jones function', and the cutoff values are $\rc=2.5$, $\tc=1$ and $\pc=\pi$ throughout this work. Throughout this work, lengths have units of $\sigma$, the subunit diameter, energies have units of $\kt$ and times have units of $t_0 = \sigma^2/D$, where $D$ is the subunit diffusion constant.  Concentrations are defined as $c_0=N \sigma^3/L^3$ with $N$ subunits and box side length $L$.

\ssect{ \change{Subunit conformations and quasi-equivalence violations.}{Conformation dependence of subunit-subunit binding energies.}} We follow convention \change{}{\cite{Johnson1997}} by labeling the different protein (monomer) conformations found in the \change{}{BMV or CCMV} crystal structure as A, B, and C. \change{The dimer subunits can have the following conformations: CC, AB, and AA, with 30 CC and 60 AB subunits comprising a $T$=3 capsid, and 30 AA subunits comprising a $T$=1 capsid.}{A $T$=3 capsid is comprised of 30 CC and 60 AB dimer subunits, while 30 AA dimer subunits comprise a $T$=1 capsid; the structure of an A protein monomer visible in the $T$=3 CCMV capsid crystal structure is virtually identical to the structure of monomers in the $T$=1 crystal structure\cite{Jinhua-note}.} For simplicity, we assume that dimer configurations not present in the crystal structures (e.g. AC, BB) are highly unfavorable and thus do not occur.

\change{According to quasi-equivalence}{If the subunit binding interactions have strict conformational specificity}, \change{a particular binding interface $b$ on subunit $i$, can bind to the complementary interface on subunit $j$ ONLY if $i$ and $j$ have the appropriate quasi-equivalent conformations}{then subunits $i$ and $j$ can only bind via a particular interface $b$ ONLY if the $i$ and $j$ have conformational states that interact via that interface in the crystal structure} (see \ref{models}). As noted above, however, there is not strong evidence to support such strict conformational specificity.  Therefore, we take $\chi_{ij}^b=1$ (Eq.~\ref{Uatt}) if the conformations of subunits $i$ and $j$ are found in the crystal structure for interface $b$, and $\chi_{ij}^b=\xs$ otherwise, where \change{}{the promiscuity parameter} $\xs$ varies from 0 for strict conformational specificity to 1 for no conformational specificity at all. \change{no penalty for violating quasi-equivalence).}{}

To capture the ability of BMV to form both $T$=1 and $T$=3 structures, we allow subunits to change conformation during the simulation. For simplicity, we  model the transitions as discrete events with no intermediate states, implying that subunit conformations are separated by an activation barrier. These moves are accepted according to the  Metropolis condition:
\begin{equation}
\mathrm{P} = \mathrm{min}\{1,\exp( - \Delta U - \Delta \varepsilon)\}
\label{eq:metropolis}
\end{equation}
where $\Delta U$ is the change in the interaction energy and $\Delta \varepsilon$ is the intrinsic free energy difference between conformational states, which might correspond to the energies associated with different hinge angles described in Ref.~\cite{Tang2006} (see \ref{hinge}). We set the energy of AB 0 and the CC energy to $\log2$ \change{due to symmetry factors}{(because it is symmetric)}. We vary the remaining energy, $\eaa$, between 0 and 2.5. \change{In contrast to the model in Ref. \cite{Nguyen2008}, the same subunits can form $T$=1 or $T$=3  capsids, due to their ability to switch conformations.}{}  For simplicity, we do not consider \emph{pseudo}T=2 capsids in this work, which unlike $T$=1 and $T$=3  capsids, involve binding interfaces that are not seen in infectious CCMV viruses\cite{Tang2006}.

\ssect { Core-controlled assembly.}

Motivated by recent experiments in which BMV capsid proteins encapsidate functionalized gold nanoparticles \cite{Dragnea2003}, we introduce a rigid sphere with radius $\rs$ at the center of the simulation. The sphere interacts with the subunits via a spherically symmetric Lennard-Jones potential, shifted so that a subunit at the surface of the sphere has minimum energy
\begin{equation}
U_S(r) = \esph \Theta(r_{\mathrm{eff}}  - r_c) \left( \scL(r_\mathrm{eff} ) - \scL(r_c)\right)
\label{eq:sphereU}
\end{equation}
where $r_{\mathrm{eff}} \equiv r - R_s$ with $r$ the nanoparticle-subunit center-to-center distance, and $\esph$ specifies the strength of the subunit-sphere interaction; we consider the range $\esph/\kt \in [6,12]$.  We consider nanoparticles with $\rs=1.7$ and $\rs=3.2$, which are commensurate with $T$=1 and $T$=3 model capsids, respectively.

\ssect { Dynamics simulations.} We evolve particle positions and orientations with over-damped Brownian dynamics using a second order predictor-corrector algorithm\cite{Branka1999, *Heyes2000}. We intersperse conformational Monte-Carlo moves with dynamics integration steps such that, on average, each particle attempts to change conformation with frequency \footnote{Assembly behavior appears to be largely independent of $\gamma_c$, except at extreme values: assembly is not adaptable at $\gamc=0$, and assembly is not productive at $\gamc=\infty$} $\gamc=(40 t_0)^{-1}$.  When there is a nanoparticle present we simulate a bulk solution with concentration $c_0$ by performing grand canonical Monte Carlo moves in which subunits far from the nanoparticle are exchanged with a reservoir at fixed chemical potential with frequency consistent with the diffusion limited rate\cite{Hagan2008}. \change{}{Since only a single nanoparticle is considered in each simulation, interactions between nanoparticles are not considered -- finite nanoparticle concentrations are considered in Ref \cite{Hagan2008b}.} Empty capsid simulations have $N=1000$ subunits in a box with side length $L=22.5 \sigma$.\newline

\begin{figure}
\centering
\includegraphics[width= \textwidth]{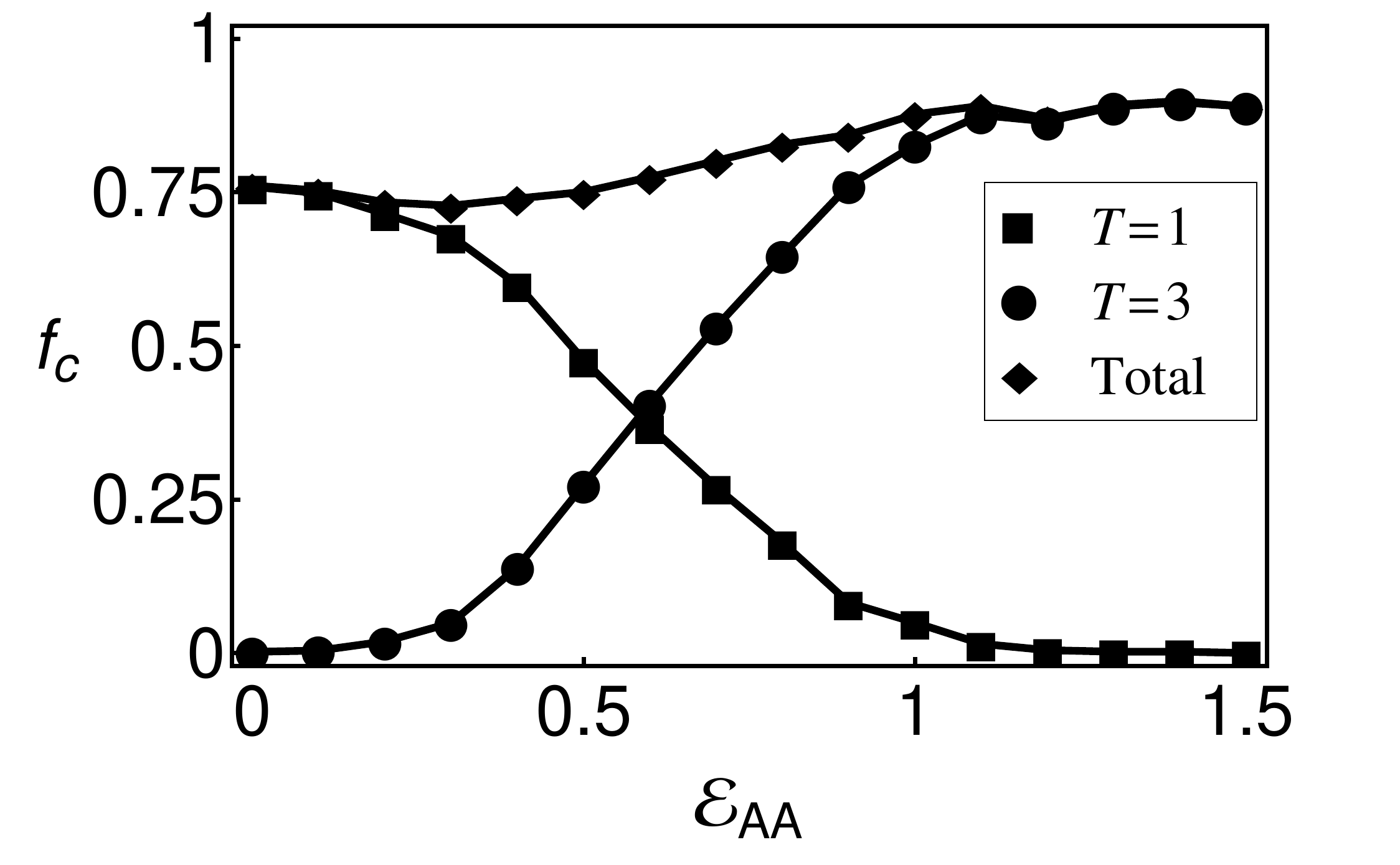}
\caption{The fraction of subunits in well-formed $T$=1 ($\blacksquare$),$T$=3 ({\Large $\bullet$}) or both ($\blacklozenge$) capsids for varying conformational free energy, $\eaa$, in empty capsid simulations.}
\label{empty_crossover}
\end{figure}

\sect {Results}
\bssect { Empty capsids.} In order to understand which system parameters control assembly morphology, we performed dynamical empty capsid simulations for varying values of the conformational free energy, $\eaa$, and the conformation specificity parameter, $\xs$, \change{which controls the binding energy for subunit-subunit bonds that violate quasi-equivalence.}{ which controls the conformational dependence of subunit-subunit binding energies.}  Prior work\cite{Hagan2006} has shown that assembly yields are nonmonotonic in the parameters that control the driving forces for assembly, the binding energy, $\eb$, and the angular tolerance, $\theta_c$. \change{In}{For empty capsid simulations in} this work, we consider only optimal values of these parameters, $\eb=12.0$,\ $\theta_c = 1.0$, for which subunits that are capable of forming only a single morphology assemble with high fidelity \cite{Hagan2008}. 

\ssect {The conformational free energy controls polymorphism in empty capsid assembly.} As a measure of  morphology control, we monitor the fraction of subunits in $T$=1 and $T$=3 capsids, which are defined as connected clusters comprising 30 or 90 subunits, respectively, in which each subunit has 4 bonds. The yield of each morphology as a function of the conformational free energy, $\eaa$, is shown in  ~\ref{empty_crossover} for an observation time of $200,000 t_0$,  at which point assembly has roughly saturated. We observe a crossover from  high yields of $T$=1 capsids for $\eaa < 0$ to  predominantly  $T$=3  capsids for $\eaa > 1.0$, with mixed morphologies in the intermediate region.

Although the transition between $T$=1 and $T$=3 capsids only requires a change in the conformational free energy of approximately the thermal energy, $k_B T$, the width of the transition is consistent with kinetic rather than thermodynamic control of the dominant morphology.  As shown in Fig 6b of the SI, fitting the fraction of subunits in $T$=1 capsids, $\ftone$ to the form $\ftone(\eaa)=1/(1+ 3 \log2  \exp[-\nmorph (\eaa - \log 2)])$, yields a `critical' size for morphology determination of $N_c \in \left[6.5, 8.0\right]$, while an equilibrated system would give a much sharper transition with $\nmorph=30$. \change{Consistent with this result, analysis of commitor probabilities as described in the SI yields a critical assembly nucleus, the minimum size of an intermediate such that the probability of growing to completion exceeds that of dissociating into monomers, of approximately 7.}{}

\change{}{To better understand this result, we estimated the size of a ``critical nucleus'', or an intermediate which is more likely to grow into a complete capsid then to dissociate into free subunits.  A simulation with 10,000 subunits was run until many small assembly intermediates (2-12 subunits) assembled. From that configuration, further simulations were integrated with different random number seeds, and each initial intermediate was tracked in every simulation.  We then estimate the ``commitor probability'' for each intermediate as the fraction of simulations in which it grows to completion before dissociating.  The average commitor probability as a function of intermediate size (figure 6a in the SI) suggests that the critical nucleus is 7, which is consistent with the critical morphology size estimated above. We note, however, that this value is only a rough estimate, since the identity of critical nuclei depends on additional parameters, such as the number of bonds and closed polygons -- i.e. the intermediate size alone is not sufficient for a good reaction coordinate.}

\begin{figure}
\centering
\includegraphics[width= \textwidth]{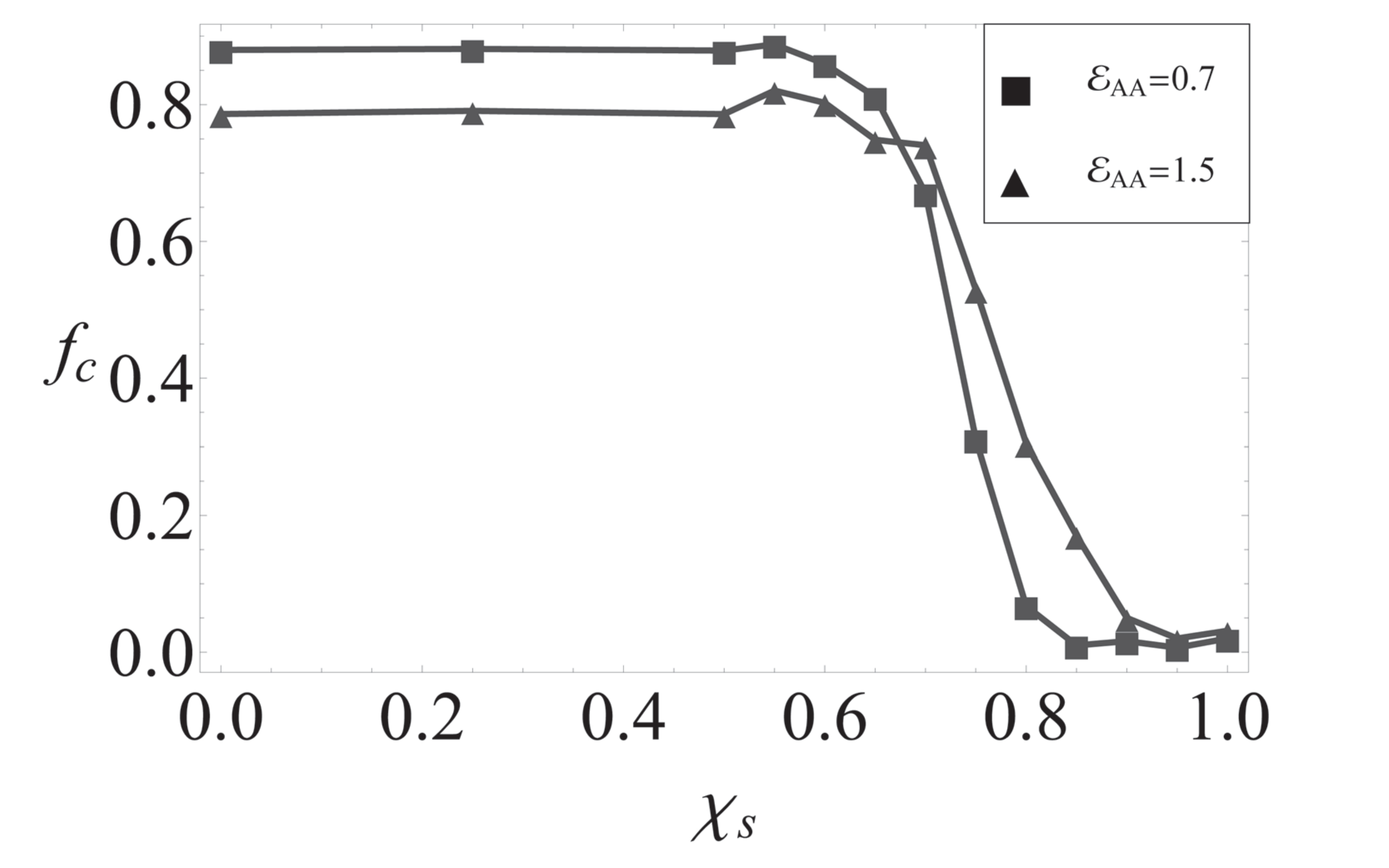}
\caption{The fraction of subunits in well-formed capsids as a function of the conformational specificity parameter $\xs$.}
\label{local_rules}
\end{figure}

\ssect { Faithful assembly requires only weak conformational specificity.} As shown in \ref{empty_crossover}, varying the intrinsic conformational free energy, $\eaa$, leads to different morphologies, but does not significantly affect the yields of well-formed icosahedrons.  As shown in \ref{local_rules}, assembly yields are also insensitive to the conformation specificity for $\xs \lesssim 75\%$, while higher values (indicating less specificity) lead to predominantly malformed capsids. These malformed capsids are primarily closed but strained structures with disordered arrangements of pentamers and hexamers that do not have icosahedral symmetry (see Figs. 2a and 2b in the SI).

\begin{figure}[!h]
\begin{center}
\subfloat[]{
\includegraphics[width=\textwidth]{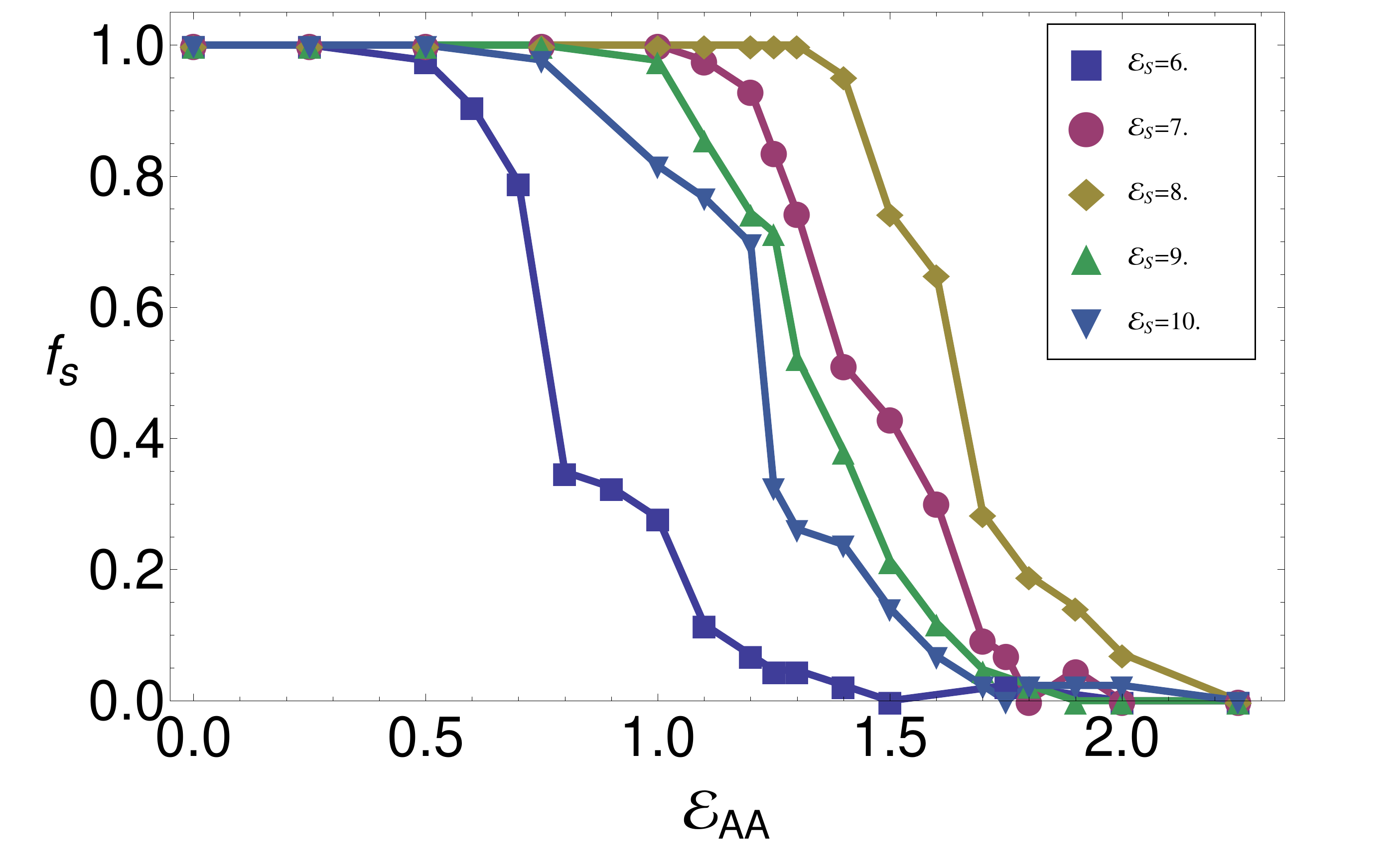}
\label{sphere_crossover}
} \\
\subfloat[]{
\includegraphics[width=\textwidth]{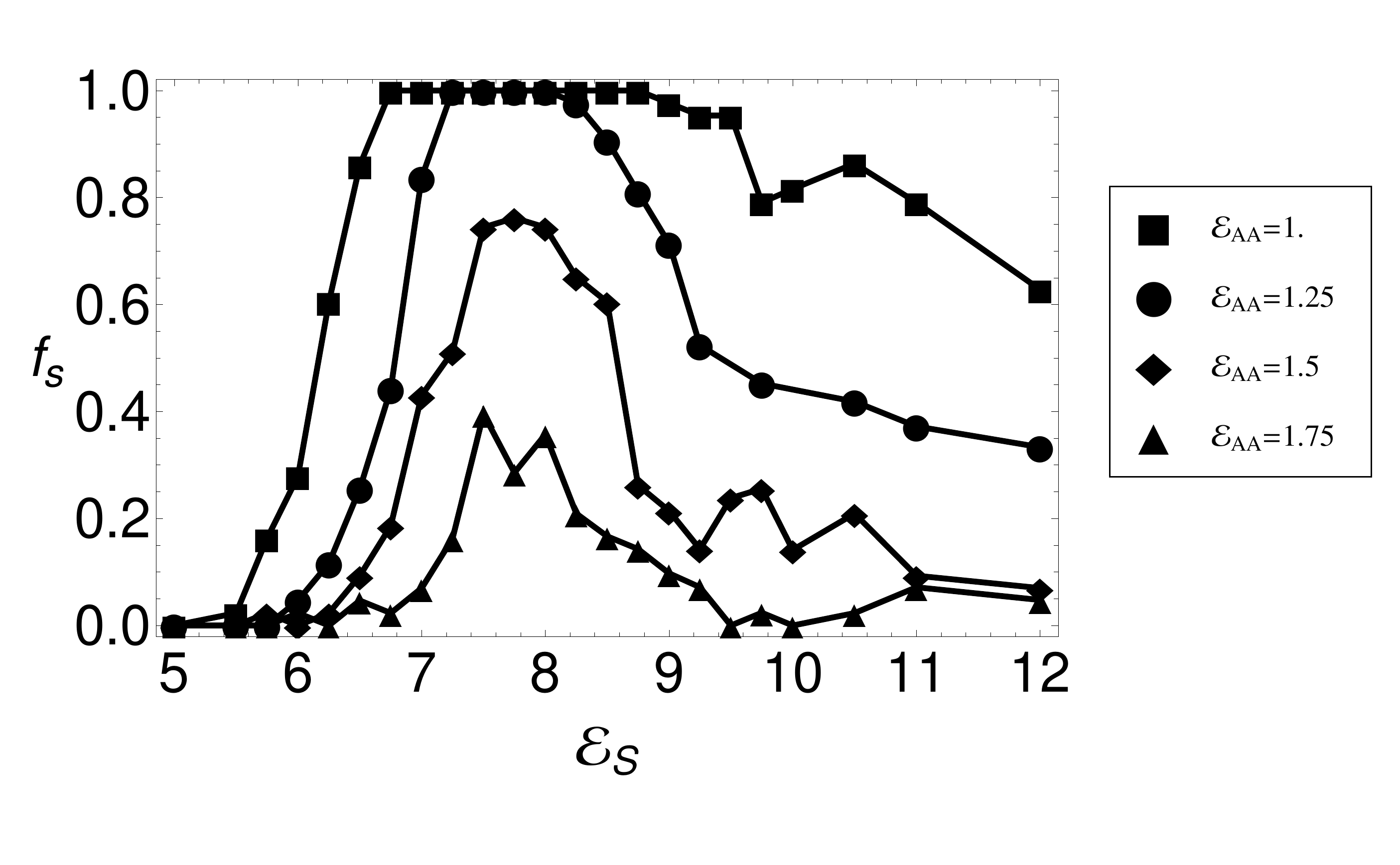}
\label{nonmonotonic}
}
\caption{The fraction of trajectories in which a complete capsid assembles on a $T$=1-size nanoparticle as functions of {\bf (a)} $\eaa$ for several $\esph$ and {\bf (b)} $\esph$ at several $\eaa$, 
All packaging efficiencies in this work are shown for an observation time of $40,000 t_0$ 
which is 100 times the average assembly time under optimal conditions
.}\label{sphere}
\end{center}
\end{figure}

\bssect { Core-controlled assembly.} To understand the effect of cargo properties on morphology, we simulated assembly in the presence of a model nanoparticle with varying conformational free energy, $\eaa$, and core-subunit attraction strength, $\es$. \change{}{Optimal conditions for the nanoparticle encapsidation experiments have higher pH and thus weaker subunit-subunit binding interactions than empty capsid experiments \cite{Dragnea2008}. Therefore, when simulating assembly with a model nanoparticle, we} consider subunit-subunit binding parameters of $\eb=10.0$,\ $\theta_c = 1.0$ for which assembly is favorable on the nanoparticle, but no spontaneous assembly in bulk solution occurs \cite{Hagan2008}.  As a measure of assembly effectiveness, we monitor the packaging efficiency, $f_\text{s}$, which is defined as the fraction of independent trajectories in which a well-formed capsid assembles on the nanoparticle. Consistent with prior work \cite{Hagan2008}, parameter values that lead to a single morphology of empty capsids enable efficient encapsidation of a commensurate sphere with $\approx 100\%$ efficiency. To understand polymorphism, we monitor packaging efficiencies around a $T$=1 sized sphere while varying the conformational free energy from values that favor $T$=1 empty capsids ($\eaa \lesssim 0.25$) to values favoring $T$=3 empty capsids ($\eaa \gtrsim 1$).  The results are shown in ~\ref{sphere_crossover} for several core-subunit interaction strengths. At each $\es$, there is a relatively sharp crossover from high yields to no successful assembly.  Significantly, for optimal values of $\es$, there is a range of $1\lesssim \eaa \lesssim1.5$ for which spontaneous assembly faithfully produces $T$=3 capsids, but $T$=1 capsids form on the nanoparticle with high efficiency.

\begin{figure}
\begin{center}
\subfloat[]{
\includegraphics[width=.45 \textwidth]{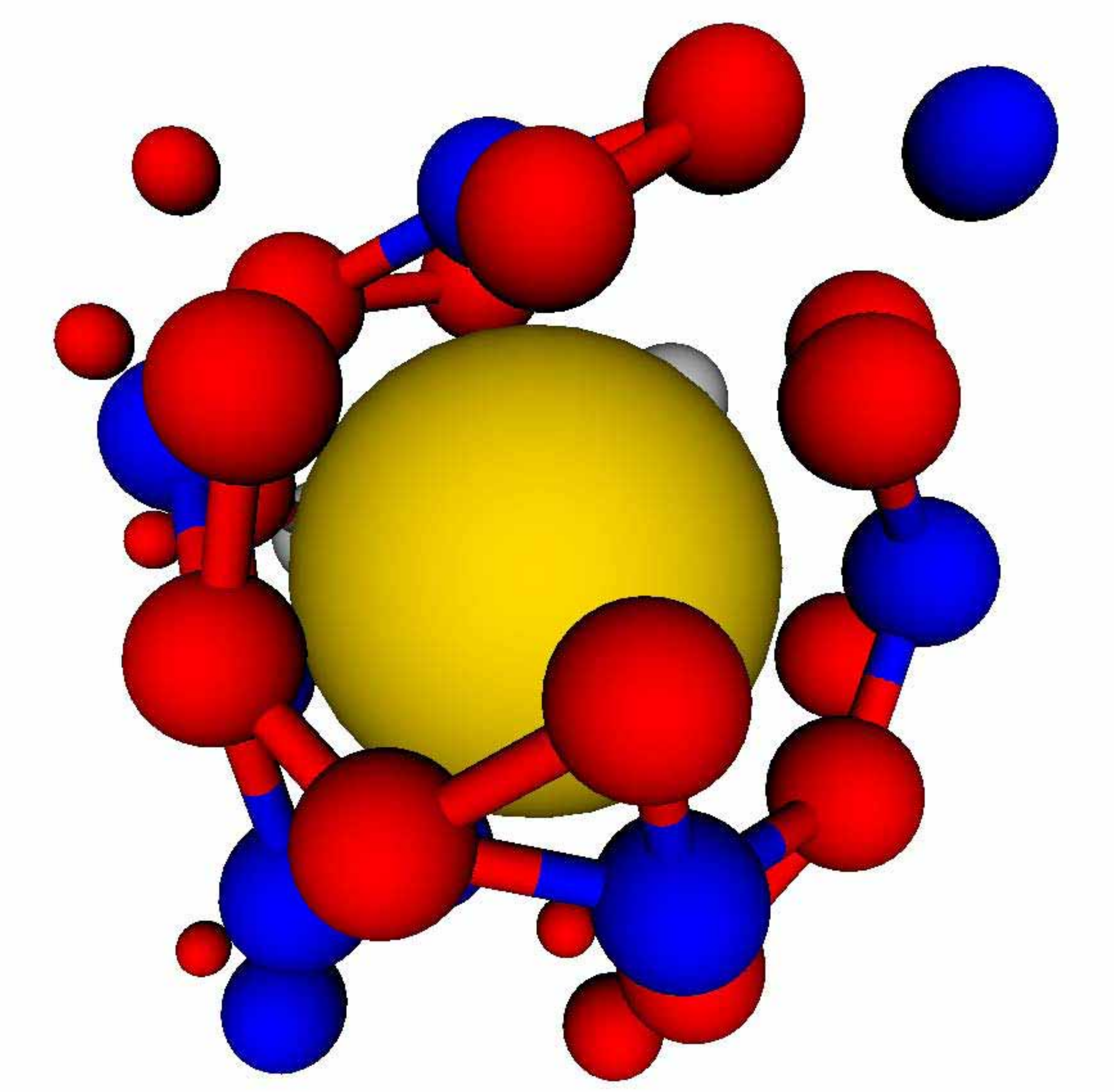}
\label{frustrated}
}
\subfloat[]{
\includegraphics[width=.45 \textwidth]{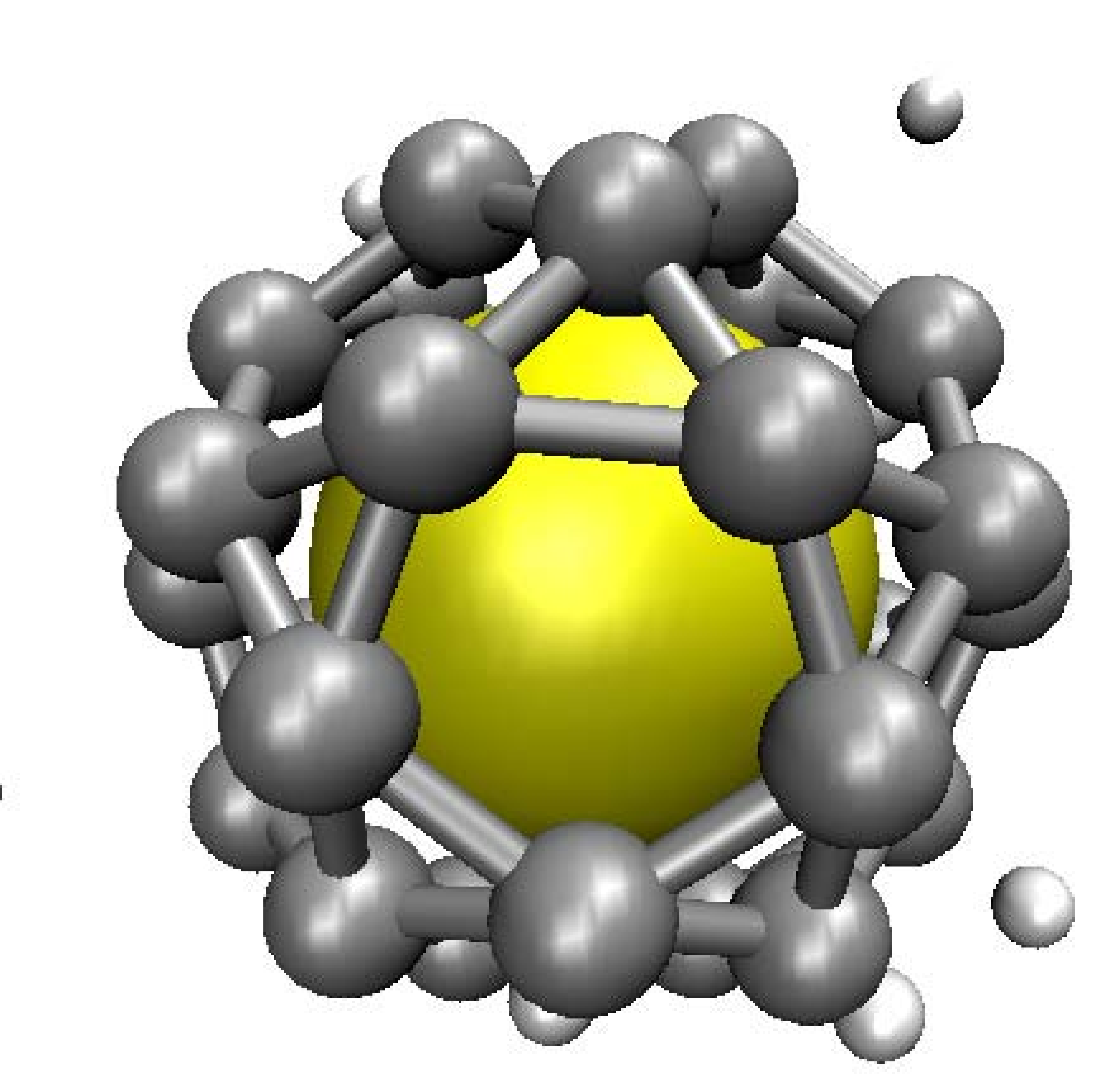}
\label{complete}
}
\caption{ {\bf a)} A partial  $T$=3  capsid grows on a $T$=1-sized nanoparticle; the subunits near the sphere are stabilized by its attraction but the mismatch between the capsid and nanoparticle geometry forbids the addition of new particles. {\bf b)} After some time, the metastable partial capsid disassembles and a $T$=1 capsid grows to completion. An animation of the entire process is available in the SI.}
\label{animate}
\end{center}
\end{figure}

\ssect {Mechanisms of core-controlled polymorphism.} In contrast to assembly around a commensurate sphere \cite{Hagan2008}, packaging efficiencies do not increase monotonically with core-subunit interaction strength ($\es$), as shown for several values of $\eaa$ in \ref{nonmonotonic}. Assembly yields that are nonmonotonic with the variation of an interaction parameter are a hallmark of competition between thermodynamics and kinetics. At low $\es$, the core-subunit interaction strength is not large enough to stabilize a $T$=1 capsid nucleus and so $T$=1 assembly is either thermodynamically unfavorable or has an insurmountable activation barrier. At high $\es$, on the other hand, capsid nuclei of any morphology are stabile and $T$=3 partial capsids are common. Beyond a certain size, subunits that add to an incommensurate ($T$=3) partial capsid cannot simultaneously interact strongly with the core surface and subunits already in the intermediate (see \ref{frustrated}).  For the parameters considered, subunit-subunit interactions are too weak to drive significant assembly away from the attractive core surface, and assembly is frustrated. At high $\es$, however, the partial capsid is metastable and blocks a significant fraction of the core surface, thus hindering the formation of potentially productive $T$=1 nuclei.

At optimal strengths of the core-subunit interaction, partial capsids undergo significant size fluctuations because subunit unbinding is frequent.  Therefore, many nuclei can form on a given sphere within the observation time, until a $T$=1 nucleus grows to completion.  ~\ref{frustrated} shows a metastable partial capsid forming on an incommensurate size sphere which eventually disassembles, allowing a $T$=1 capsid to grow to completion (\ref{complete}).

\begin{figure}
\centering
\includegraphics[width=\textwidth]{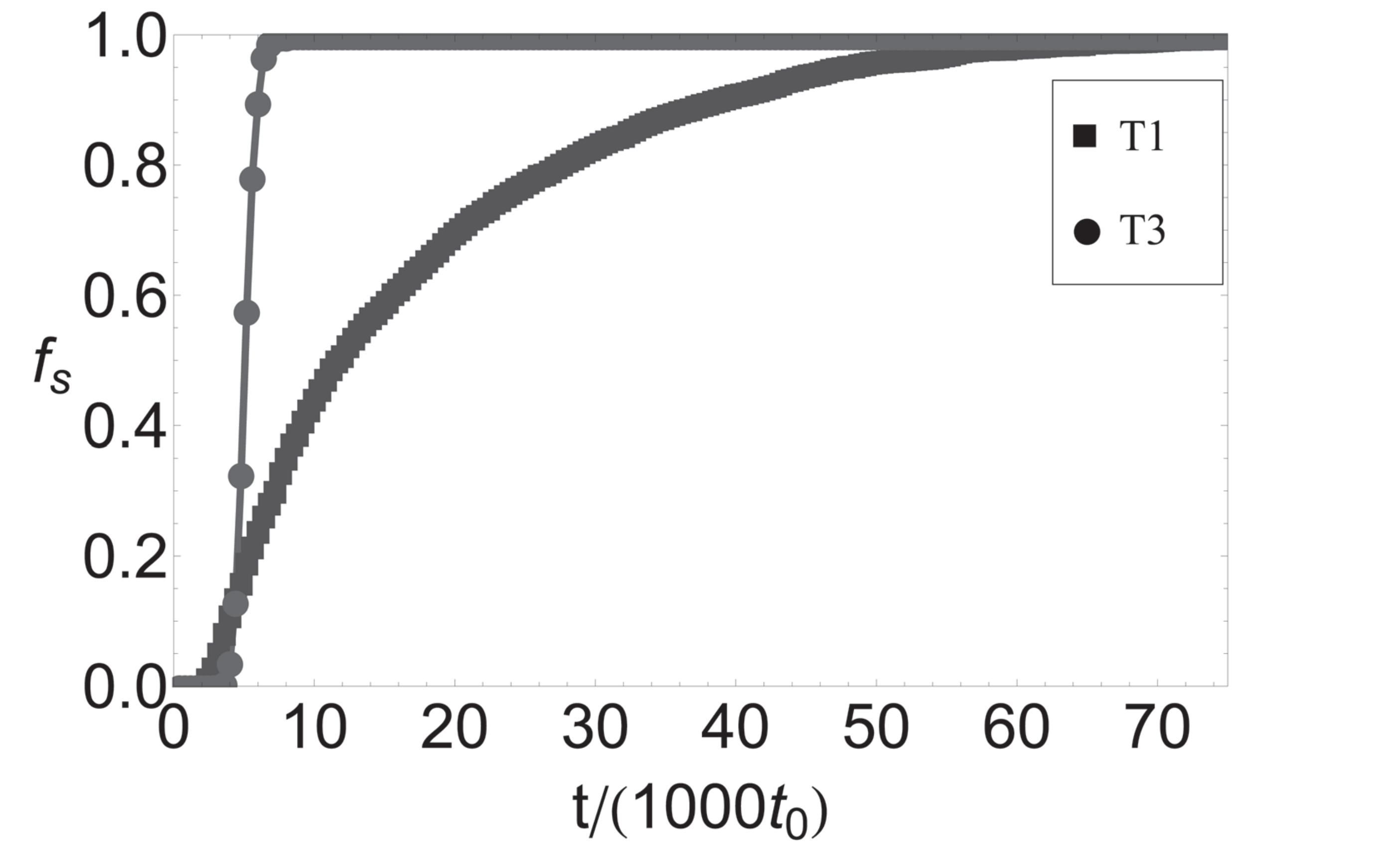}
\caption{Packaging efficiencies as functions of time for encapsidation of $T$=1-size and $T$=3-size nanoparticles for parameters at which $T$=3 capsids are the lowest free energy empty capsid structures, $\eaa=1.3$.  Other parameters are $\esph=$8.0 and $c_0=0.01$.}
\label{Yield_Times}
\end{figure}

\begin{figure}
\centering
\includegraphics[width= \textwidth]{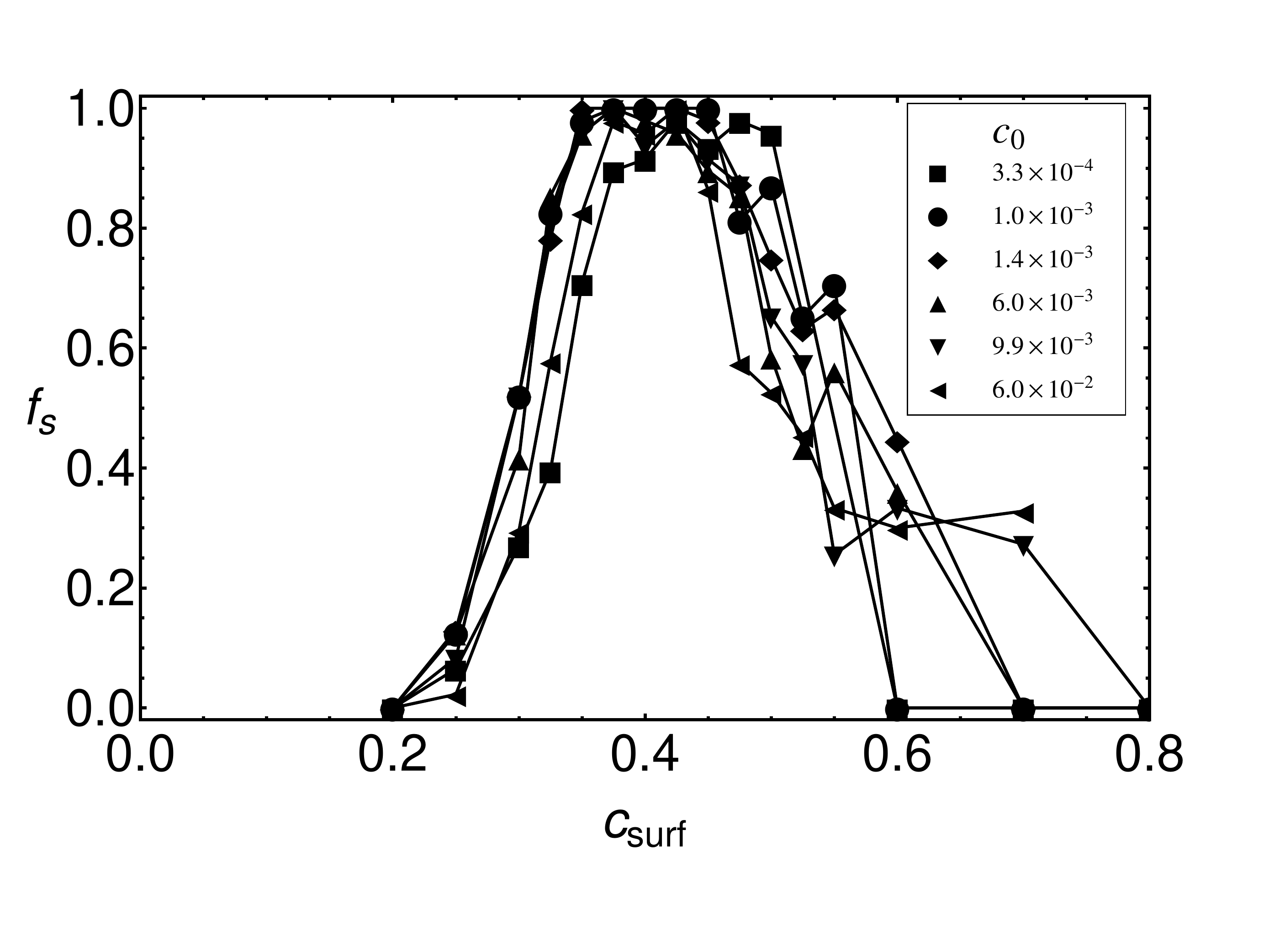}
\caption{Packaging efficiencies on a $T$=1-size nanoparticle for several bulk subunit concentrations, $c_0$ are shown for varying subunit-surface energies, plotted as functions of the surface concentration at which subunit adsorption saturates, $\csurf$.  The conformational free energy is $\eaa=1.3$.}
\label{csurf}
\end{figure}

\ssect {Time scale for annealing of surface-adsorbed complexes.} We simulated assembly with a $T$=1-size nanoparticle over a wide range of subunit concentrations, $3 \times 10^{-4} < c_0 < 0.06$, which corresponds to a range of $7-1000 \mu M$ for a dimer subunit diameter of 4.2 nm. The analysis in Ref. \cite{Hagan2008} suggests that nanoparticle systems with different $\esph$ and $c_0$ should be compared in terms of $\csurf$,  the equilibrium surface concentration of subunits with no assembly\protect\footnote{ $\csurf \approx  \frac{n_s}{4 \pi R_s^2}$ where $n_s$ is the number of subunits with strong ($<-2\kt$) interactions with the nanoparticle.}. The equilibrium surface concentration can be determined from simulations with $\eb = 0$ or by calculating the chemical potential of adsorbed subunits (see the SI). The packaging efficiencies for various $\esph$ and $c_0$ are plotted as a function of $\csurf$ in \ref{csurf}. We see that optimal assembly for all concentrations collapses onto the same value of $\csurf \approx 0.4$, while higher values of $\csurf$ \change{meanly}{mainly} lead to trapped incommensurate partial capsids.  Interestingly, successful assembly occurs at higher $\csurf$ if the rates of subunit adsorption are decreased below the diffusion limited rate by decreasing the frequency of subunit exchanges in the bath (see Figure 6 in the SI), when subunits adsorb more slowly compared to the timescale for annealing of surface-adsorbed complexes.  We therefore note that effects which increase the surface-annealing timescale, such as barriers to diffusion of strongly adsorbed subunits, could further promote frustrated states.

\clearpage

\ssect {Larger Capsid Morphologies}

\change{}{To further explore the requirements for adaptability to template size, we extended our model to include $T$=4 capsids, which requires at least one new subunit geometry in addition to AB and CC (see page 3 of the SI). We note, however, that Sun et al.\cite{Sun2007} observe only disordered structures on $T$=4-size nanoparticles, which may suggest $T$=4 subunit geometries are not favorable for BMV proteins.  As shown in figure 4 in the SI, these model subunits are capable of forming $T$=3 or $T$=4 capsids around nanoparticles ranging from $T$=3 to $T$=4 size; however, yields of perfect capsids are generally lower ($\lessim 50\%$) and more sensitive to parameter values than for the $T$=1/$T$=3 case.  In addition to intermediates that cannot close around an incommensurate core (as described above), there are significant yields of closed but disordered asymmetric  shells.  This result is consistent with the fact that the difference in curvature between $T$=3 and $T$=4-size nanoparticles is small ($\approx 12\%$) and thus templating for the commensurate capsid geometry is weak, especially for small intermediates.}

\sect { Discussion}
\change{\ssect { Parameter values in experimental systems.}}{\ssect {Estimating the conformational free energy from experiments.}} Comparison of model predictions to experimental observations of capsid morphology suggests a potential correspondence between ranges of the conformational free energy, $\eaa$, and certain systems. In particular, CCMV and BMV capsid proteins assemble into exclusively $T$=3 empty capsids under conditions commonly employed in vitro, while deletion of the N-terminal residues from the proteins of either virus enables assembly of $T$=1 capsids\cite{Tang2006}. In particular, proteins of the $\nd$ CCMV mutant, with 34 N-terminal residues deleted, assemble into a mix of $T$=3 ($\approx 10\%$ of assembled material), $T$=1 ($\approx 40\%$) and heterogeneous \emph{pseudo}T2 assemblages ($\approx 50\%$), could correspond to the range $0.3 \lessim \eaa \lessim 0.8$ for which polymorphism is observed.  We note that this range could shift somewhat, however, if there are different binding energies for different complementary interfaces and because we have not considered \emph{pseudo}T2 capsids in this work, since they require interfacial contacts that are not seen in the $T$=3 crystal structure (unlike $T$=1 capsids) \cite{Tang2006}.

The observation that wild-type CCMV and BMV proteins form exclusively $T$=3  capsids can only suggest a lower bound for the wild-type conformational free energy, $\eaa \greatsim 1.2$, but additionally considering nanoparticle experiments allows a more precise estimate. Comparison of Figs.~\ref{empty_crossover} and \ref{nonmonotonic} identifies only a narrow range $1.1 \lessim \eaa \lessim 1.5$ for which predominantly $T$=3 empty capsids form, but $T$=1 capsids efficiently encapsidate $T$=1-sized nanoparticles, as observed in BMV-nanoparticle experiments\cite{Dragnea2003}.  Although CCMV proteins have not been used in nanoparticle experiments,  pseudo-T2, $T$=3, and larger (but asymmetric) CCMV capsids assemble around inorganic polyelectrolytes\cite{Hu2008} and nanoemulsion droplets \cite{Chang2008}.

\change{}{Although we have focused on the relationship between nanoparticle surface charge and assembly effectiveness, other model parameters can be varied in experiments as well. For instance, decreasing pH (or increasing ionic strength) can increase subunit binding free energies\cite{Kegel2004} ($\eb$) while simultaneously decreasing the subunit-nanoparticle interaction strength ($\es$), since the nanoparticle surfaces are functionalized with carboxylated PEG. Thus, the optimal pH for encapsulation of nanoparticles is larger than the optimal pH for empty capsid assembly \cite{Dragnea2008}. Therefore we have simulated assembly in the presence of nanoparticles with a subunit binding energy of $\eb=10$, which is lower than the optimal subunit binding energy $\eb=12$ for empty capsid assembly; the interdependence of assembly effectiveness on subunit-subunit and the subunit-nanoparticle interaction strengths is explored in Ref \citep{Hagan2008b}.  We also note that the conformation free energy (hinge energy) could depend on pH.} 

\ssect {Implications for quasi-equivalence.} The predictions of our model many shed light on the mechanisms by which subunits can assemble with precise spatial ordering of different conformations even though the bonding interfaces in different conformations are structurally similar. In particular, the model predicts that assembly products, and to some extent assembly times, are insensitive to the inter-subunit conformational binding specificity, $\xs$, for $\xs \greatsim 25\%$. For the parameters used in this work, the free energy per bond in a complete capsid is approximately 3.5 $k_B T$ (this estimate includes the entropy penalty for the subunit binding, see \cite{Hagan2006}). Thus successful assembly requires only that conformational pairings \change{that violate quasi-equivalence}{not seen in the crystal structure} differ by $\approx k_B T$ from native pairings, which could arise from only minor variations in binding interfaces. Because optimal assembly occurs for weak subunit-subunit interactions, when subunit binding is only slightly more favorable than unbinding \cite{Ceres2002, *Jack2007, *Rapaport2008}, a small difference in subunit binding free energies, combined with the strain caused by the geometrical incompatibilities that result from non-native bonding, strongly favors a Caspar-Klug capsid structure.

\ssect { Suggested experiments.}  A key prediction of our work is that assembly on a nanoparticle with a size that does not match the lowest free energy capsid morphology is impeded by partial capsids whose curvature is inconsistent with the nanoparticle surface. These frustrated states are revealed in several ways that may be accessible to experiments. First, simulated packaging efficiencies on $T$=1-size nanoparticles, for subunits that form $T$=3 empty capsids, are nonmonotonic with respect to the nanoparticle-subunit interaction strength (\ref{nonmonotonic}).  This parameter is controlled in experiments by functionalizing nanoparticle surfaces with different ratios of anionic and neutral molecules \cite{Dragnea2008}. In Ref. \cite{Hagan2008} we show that the chemical potential of adsorbed subunits, and hence the equilibrium surface concentration $\csurf$ (see \ref{csurf}), can be estimated from the surface density of functionalized charge; $\csurf$ can be varied over the range $0\le \csurf \lesssim 0.8$.  \change{}{Although the degree of nonmonotonicity depends on factors such as subunit adsorption rates and surface annealing rates}, frustration can still be observed by differences in assembly kinetics on $T$=1 and $T$=3-size nanoparticles even for conditions in which eventual packaging efficiencies reach 100\% as in \ref{Yield_Times}.

\sect {Summary.}
\change{In summary, we}{We} have performed simulations with a model for assembly of subunits into empty capsids and around nanoparticles that template the assembly of different morphologies.  The simulations uncover mechanisms by which assembly can adapt to form different morphologies when challenged with a template that does not match the preferred empty capsid structure.  Predicted assembly pathways include frustrated partial capsid intermediates with curvatures that do not match the template, which leads to predicted differences in assembly kinetics and effectiveness on nanoparticles with different sizes, that can be tested in experiments.  These findings may shed light on the role of nucleic acids in assembly during viral replication, and demonstrate that the interplay between the geometries of different components is an important consideration for the design of nanostructured materials.

\sect{Outlook.}
\change{}{Extensions to this study could include an explicit representation of intra-subunit degrees of freedom (i.e. hinge motions of the protein dimer), and template degrees of freedom. As suggested by an anonymous reviewer, a flexible spherical template could represent a nucleic acid molecule with significant structure due to base pairing. We find that assembly around a flexible spherical template has qualitatively similar results to those reported here (for some ranges of sphere flexibility), although the optimal subunit-template interaction strength increases somewhat; these results will be presented in a future publication.}

In order to focus on the effects of template-capsid geometry mismatches on assembly, we have not considered other potential sources of frustration, such as impeded diffusion for subunits that interact strongly with the nanoparticle surface.  The coupling of multiple sources of frustration could have interesting consequences.

\sect{Supplementary Materials}

Supplementary materials are available \href{http://people.brandeis.edu/~elrad/poly_paper_si.pdf}{online.}

\sect{Acknowledgments}

\change{}{We gratefully acknowledge Chris Henley for insightful comments and Jinghua Tang for providing the $T$=1 CCMV crystal structure.} Funding was provided by an HHMI-NIBIB Interfaces Initiative grant to Brandeis University and Brandeis University startup funds. \change{}{Simulations were performed on the High Performance Computing Cluster at Brandeis University.}

\footnotesize
\single
\bibliographystyle{achemsoM}

\ifx\mcitethebibliography\mciteundefinedmacro
\PackageError{achemso.bst}{mciteplus.sty has not been loaded}
{This bibstyle requires the use of the mciteplus package.}\fi

\end{document}